\documentclass[12pt,a4paper]{article}

\usepackage{amsmath}
\usepackage{amsthm}
\usepackage{amssymb}
\usepackage[left=1cm,right=1cm,top=2cm,bottom=2cm]{geometry}

\def\be{\begin{eqnarray}}
\def\ee{\end{eqnarray}}

\begin{document}
\hfill ITEP/TH-36/12

\bigskip
\centerline{\Large{Special colored Superpolynomials and their representation-dependence}}
\bigskip
\centerline{Anton Morozov\footnote[1]{anton.morozov@itep.ru}}

\bigskip

\centerline{{\it Moscow State University and ITEP, Moscow, Russia}}

\bigskip

\centerline {ABSTRACT}
\bigskip
{\footnotesize
We introduce the notion of "special superpolynomials" by putting $\ q=1\ $ in the formulas for reduced superpolynomials. In this way we obtain a generalization of special HOMFLY polynomials depending on one extra parameter $t$. Special HOMFLY are known to depend on representation $R$ in especially simple way: as $|R|$-th power of the fundamental ones. We show that the same dependence persists for our special superpolynomials in the case of symmetric representations, at least for the 2-strand torus and figure-eight knots. For antisymmetric representations the same is true, but for $t=1$ and arbitrary $q$. It would be interesting to find an interpolation between these two relations for arbitrary representations, but no superpolynomails are yet available in this case.
}
\bigskip

\section{Introduction}

Colored superpolynomials \cite{superpolsfirst} become increasingly important objects in the fields of knot theory \cite{CS} and  topological strings \cite {Vetco}. They are functions of five variables: knot/link $K$, representation $R$ and three parameters $\bf{q}$, $\bf{a}$ and $\bf{t}$. For our purposes it is more convenient to use other variables \cite{DMMSS}: \be q=-{\bf tq},\ \ \ \ t={\bf q},\ \ \ \ A={\bf a}\sqrt{-{\bf t}}\  \label{1} \ee In these variables one can study nontrivial limits, obscure in original notation. HOMFLY polynomials arise at $t=q$. If further $t=q=1$ we obtain the special HOMFLY polynomials which posses remarkable property \cite{DMMSS,IMMM}:\footnote{It is important that HOMFLY polynomials are reduced: otherwise the limit does not exist.}
\be
\boxed{{\cal H}^K_R(A)=\left({{\cal H}^K_\square  (A)}\right)^{|R|}} \label{4}
\ee where $\square$ is used to denote the fundamental representation and $|R|$ is the number of boxes in the Young  diagram $R$. For rigorous proof of (2), also in the case of links, see \cite{zhu}.

It is  natural to ask what happens to this relation when HOMFLY is lifted to superpolynomial. We conjecture that it turns into
\be
\boxed
{{\cal P}^{K}_{S^r}(A,t)=\left({\cal P}^{K}_{\square}(A,t)\right)^r} \label{2}
\ee
for arbitrary symmetric representation $S^r$ and
\be
\boxed{{\frak P}^{K}_{\Lambda^r}(A,q)= \left({\frak P}^{K}_{\square}(A,q)\right)^r} \label{3}
\ee
for arbitrary antisymmetric  one $\Lambda^r$. Here $\cal{P}\ $and$\ \frak{P}$ are defined as the limits of reduced superpolynomial at $q\rightarrow1$ and $t\rightarrow1$ respectively.

We check this conjecture in all available examples: namely 2-strand torus knots and figure-eight knot. Hopefully it is true for all other knots as well.

More problematic is what happens in other representations. Not a single example is yet known of such superpolynomial and interpolation between (\ref{2}) and (\ref{3}) remains a mystery.

In what follows we  present existing evidence in support of (\ref{2}) and (\ref{3}) and conclude with a brief review of similar known relations.

\section{Some facts about particular cases of superpolynomials}
It is instructive to begin with a simple example - a trefoil. Reduced superpolynomial for it in the fundamental representation is equal to:
\\
\centerline{
$\begin{array}{ccccc}
& & P^{[2,3]}_{\square}(A,q,t) = -A^2q^2+q^2t^2+1& &  \\ \\
& \swarrow^{q\rightarrow1}\!\! & &\!\! \searrow^{t\rightarrow1} & \\ \\
{\cal P}^{T[2,3]}_{\square}(A,t) = -A^2+t^2+1 \!\!\!\!\!\!\!\!\!\!\!\!\!\!\!
& & & &
\!\!\!\!\!\!\!\!\!\!\!\!\!\!\! {\frak P}^{T[2,3]}_{\square}(A,q) = -A^2q^2+q^2+1 \\\\
& \searrow^{t\rightarrow1} \!\!\!\!\!& &\!\!\!\!\! \swarrow^{q\rightarrow1} & \\\\
& & {\cal H}^{T[2,3]}_{\square}(A)= -A^2 +2 & &\\\\
\end{array}$}
\bigskip
We showed also what happens when we put $q$ or $t$ equal one.

  Now let us do the same for trefoil superpolynomial in the first symmetric representation $[2]=S^2$, taken from \cite{DMMSS}:\\
\centerline{
$\begin{array}{ccccc}
& & P^{[2,3]}_{[2]}(A,q,t) = q^{10}A^4+(q^{10}t^2+q^8t^2+q^6+q^4)A^2+q^8t^4+q^6t^2+q^4t^2+1& &  \\\\
& \swarrow^{q\rightarrow1} \!\!\!\!\!\!\!\!\!\! \!\!\!\!\!\!\!\!\!\! \!\!\!\!\!\!\!\!\!\! & & \!\!\!\!\!\!\!\!\!\! \!\!\!\!\!\!\!\!\!\! \!\!\!\!\!\!\!\!\!\! \searrow^{t\rightarrow1} & \\\\
{\cal P}^{T[2,3]}_{[2]}(A,t) = A^4-2(t^2+1)A^2+(t^2+1)^2 = \!\!\!\!\! \!\!\!\!\! \!\!\!\!\! \!\!\!\!\!\!\!\!\!\! \!\!\!\!\!\!\!\!\!\!\!\!\!\!\!\!\!\!\!\!\!\!\!\!\!\!\!\!\!\!\!\!\!\!\!\!\!\!\!\!\!\!\!\!\!\!\!\!\!\!\!\!\!\!\!\!\!\!\!\!\!\!\!\
\!\!\!\!\!\!\!\!\!\!\!\!\!\!\!\! & & & & \!\!\!\!\!\!\!\!\!\!\!\!\!\!\!\!\!\!\!\!\!\!\!\!\!\!\!\!\!\!\!\!\!\!\!\!
\!\!\!\!\!\!\!\!\!\!\!\!\!\!\!\!\!\!\! \!\!\!\!\! \!\!\!\!\! \!\!\!\!\!\!\!\!\!\!\!\!\!\!\!
{\frak P}^{T[2,3]}_{[2]}(A,q) = q^{10}A^4-(q^{10}+q^8+q^6+q^4)A^2+\\ =\boxed{\left(-A^2+t^2+1\right)^2}\!\!\!\!\!\!\!\!\!\!\!\!\!\!\!\!\!\!\!\!\!\!\!\!\!\!\!\!\!\!\!\!\!\!\!\!\!\!\!\!\!\!\!\!\!\!\
\!\!\!\!\!\!\!\!\!\!\!\!\!\!\!\! \!\!\!\!\!\!\!\!\!\!\!\!\!\!\!\!\!\!\!\!\!\!\!\!
\!\!\!\!\!\!\!\!\!\!\!\!\!\!\!\!\!\!\!\!\!\!\!\!&&&&\!\!\!\!\!\!\!\!\!\!\!\!\!\!\!\!\!\!\!\!\!\!\!\!\!\!\!\!\!\!\!\!\!\!\!\!\!\!\!\!\!\!\!\!\!\!\
\!\!\!\!\!\!\!\!\!\!\!\!\!\!\!\! \!\!\!\!\!\!\!\!\!\!\!\!  +q^8+q^6+q^4+1\\ \\
& \searrow^{t\rightarrow1}\!\!\!\!\!\!\!\!\!\! \!\!\!\!\!\!\!\!\!\! \!\!\!\!\!\!\!\!\!\! \!\!\!\!\!\!\!\!\!\!  \!\!\!\!\!\!\!\!\!\! \!\!\!\!\!\!\!\!\!\!  & & \!\!\!\!\!\!\!\!\!\! \!\!\!\!\!\!\!\!\!\! \!\!\!\!\!\!\!\!\!\! \!\!\!\!\!\!\!\!\!\! \!\!\!\!\!\!\!\!\!\! \!\!\!\!\!\!\!\!\!\!  \swarrow^{q\rightarrow1} & \\\\
& & {\cal H}^{T[2,3]}_{[2]}(A)= A^4 -4A^2 +4 = \boxed{\left(A^2-2\right)^2} & &\\
\end{array}$} \\ \\
It's clear that not only the special HOMFLY in the last line but also one of the special superpolynomials in the middle line becomes a full square. Moreover it is a square of the same quantity in the fundamental representation. It is also clear that nothing special happens with the second special superpolynomial.

However for antisymmetric representation their roles are changed:
\\ \\ \\
\centerline{
$\begin{array}{ccccc}
& & P^{[2,3]}_{[1,1]}(A,q,t) = q^{4}A^4+(q^4t^6+q^4t^4+q^2t^2+q^2)A^2+q^4t^{10}+q^2t^6+q^2t^4+t^2& &  \\\\
& \swarrow^{q\rightarrow1} \!\!\!\!\!\!\!\!\!\! \!\!\!\!\!\!\!\!\!\! \!\!\!\!\!\!\!\!\!\! \!\!\!\!\!\!\!\!\!\! & & \!\!\!\!\!\!\!\!\!\! \!\!\!\!\!\!\!\!\!\! \!\!\!\!\!\!\!\!\!\! \searrow^{t\rightarrow1} & \\\\
{\cal P}^{T[2,3]}_{[1,1]}(A,t) = A^4-(t^6+t^4+t^2+1)A^2+ \!\!\!\!\!\!\!\!\!\! \!\!\!\!\!\!\!\!\!\! \!\!\!\!\!\!\!\!\!\! \!\!\!\!\!\!\!\!\!\! \!\!\!\!\!\!\!\!\!\!  \!\!\!\!\!\!\!\!\!\! \!\!\!\!\!\!\!\!\!\! \!\!\!\!\!\!\!\!\!\!   & & & & \!\!\!\! \!\!\!\!\!\!\!\!\!\! \!\!\!\!\!\!\!\!\!\! \!\!\!\!\!\!\!\!\!\! \!\!\!\!\!\!\!\!\!\! \!\!\!\!\!\!\!\!\!\! \!\!\!\!\!\!\!\!\!\! \!\!\!\!\!\!\!\!\!\! \!\!\!\!\!\!\!\!\!\! \!\!\!\!\!\!\!\!\!\! \!\!\!\!\!\!\!\!\!\! \!\!\!\!\!\!\!\!\!\! {\frak P}^{T[2,3]}_{[1,1]}(A,q) = q^2A^4-2q^2(q^2+1)A^2+(q^2+1)^2 = \\
+(t^{10}+t^6+t^4+t^2) \!\!\!\!\!\!\!\!\!\!\!\!\!\!\!\!\!\!\!\!\!\!\!\!\!\!\!\!\!\!\!\!\!\!\!\!\!\!\!\!\!\!\!\!\!\!\
\!\!\!\!\!\!\!\!\!\!\!\!\!\!\!\!\!\!\!\!\!\!\!\!\!\!\!\!\!\!\!\!\!\!\!\!&&&& \!\!\!\!\!\!\!\!\!\! \!\!\!\!\!\!\!\!\!\! \!\!\!\!\!\!\!\!\!\!\!\!\!\!\!\!\!\!\!\!\!\!\!\!\!\!\!\!\!\!\!\!\!\!\!\!\!\!\!\!\!\!\!\!\!\!\!\!\!\!\!\!\!\!\!\!\
\!\!\!\!\!\!\!\!\!\!\!\!\!\!\!\! \!\!\!\!\!\!\!\!\!\! = \boxed{\left(-A^2q^2+q^2+1\right)^2}  \\ \\
& \searrow^{t\rightarrow1} \!\!\!\!\!\!\!\!\!\!  \!\!\!\!\!\!\!\!\!\!  \!\!\!\!\!\!\!\!\!\!  \!\!\!\!\!\!\!\!\!\!  \!\!\!\!\!\!\!\!\!\!  \!\!\!\!\!\!\!\!\!\!  \!\!\!\!\!\!\!\!\!\!  \!\!\!\!\!\!\!\!\!\!  & &\!\!\!\!\!\!\!\!\!\!  \!\!\!\!\!\!\!\!\!\!  \!\!\!\!\!\!\!\!\!\!  \!\!\!\!\!\!\!\!\!\!  \!\!\!\!\!\!\!\!\!\!  \!\!\!\!\!\!\!\!\!\!  \!\!\!\!\!\!\!\!\!\!  \swarrow^{q\rightarrow1} & \\\\
& & {\cal H}^{T[2,3]}_{[1,1]}(A)= A^4 -4A^2 +4= \boxed{\left(A^2-2\right)^2} & &
\end{array}$}
\\ \\ \\
This observations lies in the basis of our conjecture. It is easy to see that this property survives in generalizations: as we demonstrate below, one can change both representations and knots.
\section{General proof for reduced symmetric and antisymmetric superpolynomials of $2$-strand torus knots}
These colored superpolynomials were considered in \cite{IMMM, FGS}. In this section we use general formulas
from \cite{FGS} converted to variables (\ref{1}).
These formulas are represented as q-hypergeometric sums
involving the ratios of q-Pochhammer symbols $(x,y)_n=\prod\limits^{n-1}_{k=0}(1- xy^k)$
\subsection{%${\cal P}^{T[2,2p+1]}_{S^r}(A,t)$
Symmetric representations}
According to \cite{FGS}, the reduced two-strand superpolynomial in generic symmetric representation $S^r$ is:
\be P^{T[2,2p+1]}_{S^r}=\sum\limits^r_{l=0}
\frac
{(t^2;q^2)_l(q^2;q^2)_r(A^2;q^2)_{r+l}(\frac{A^2}{t^2};q^2)_{r-l}}
{(q^2;q^2)_l(A^2;q^2)_r(q^2t^2;q^2)_{r+l}(q^2;q^2)_{r-l}}
\frac{1-q^{4l}t^2}{1-t^2}(-1)^{r^2-l^2}t^{r(2+p)-l(2+2p)}A^{2rp}q^{c(r,p)}
\ee where $c(r,p)$ is known but inessential for our purposes.

  When $q$ tends to one, the q-Pochhammer symbols $(q^2;q^2)_r$, $(q^2;q^2)_l$ and $(q^2;q^2)_{r-l}$ tend to zero, but the ratio
$\frac{(q^2;q^2)_r}{(q^2;q^2)_l (q^2;q^2)_{r-l}}$ remains finite and non-vanishing. In fact it turns into binomial coefficient $C^r_l = \frac{r!}{l!(r-l)!}$. Consequently in this limit
\be
P^{T[2,2p+1]}_{S^r}\rightarrow\sum\limits^r_{l=0} \frac
{(1-t^2)^l (1-A^2)^{r+l} (1-\frac{A^2}{t^2})^{r-l}}
{(1-A^2)^r (1-t^2)^{r+l}}
\frac{r!}{(r-l)!\,l!}
(-1)^{r^2-l^2}t^{r(2+p)-2l(1+p)}A^{2rp}
\ee
or simply
\be \lim\limits_{q \rightarrow 1} \left( P^{T[2,2p+1]}_{S^r} \right) =
(-1)^r\frac
{(t^2-A^2)^r}
{(1-t^2)^r}
t^{rp}A^{2rp}
\sum\limits_{l=0}^r
\frac
{(1-A^2)^l}
{(t^2-A^2)^l}(-1)^lC^r_lt^{-2lp}
\ee
Since $\ \sum\limits_{l=0}^r\frac{(1-A^2)^l}{(t^2-A^2)^l}(-1)^lC^r_lt^{-2lp} =
\left(1-\frac{1-A^2}{t^2-A^2}t^{-2p}\right)^r$ we obtain:
\be
\boxed
{{\cal P}^{T[2,2p+1]}_{S^r}=\lim_{q\rightarrow1}P^{T[2,2p+1]}_{S^r}=
(-1)^r\frac
{(t^2-A^2)^r}
{(1-t^2)^r}
t^{rp}A^{2rp}
\left(1-\frac{1-A^2}{t^2-A^2}t^{-2p}\right)^r=\left({\cal P}^{T[2,2p+1]}_{S^1}\right)^r}
\label{sts}
\ee
So we see that our conjecture (\ref{2}) is correct. for generic $p$ and $r$!
\subsection{${\frak P}^{T[2,2p+1]}_{\Lambda^r}(A,q)$Torus antisymmetrical superpolynomials with $t\rightarrow1$}
According to \cite{FGS}, reduced two strand superpolynomial in generic symmetric representation $\Lambda^r$ is:
\be P^{T[2,2p+1]}_{\Lambda^r}=\sum\limits^r_{l=0}
\frac
{(q^2;t^2)_l(t^2;t^2)_r(A^{-2};t^2)_{r+l}(A^{-2}q^{-2};t^2)_{r-l}}
{(t^2;t^2)_l(A^{-2};t^2)_r(t^2q^2;t^2)_{r+l}(t^2;t^2)_{r-l}}
\frac{1-t^{4l}q^2}{1-q^2}(-1)^{r+l}q^{2r-rp+2lp}A^{2r(p+1)}t^{\varsigma(r,p)}
\ee
where $\varsigma(r,p)$ is known but inessential for our purposes.

When $t$ tends to one, the q-Pochhammer symbols $(t^2;t^2)_r$, $(t^2;t^2)_l$ and $(t^2;t^2)_{r-l}$ tend to zero, but the ratio $\frac{(t^2;t^2)_r}{(t^2;t^2)_l (t^2;t^2)_{r-l}}$  remains finite and no vanishing. In fact it turns into binomial coefficient $C^r_l$ as it was in the symmetric case. Consequently in this limit:
\be
\lim\limits_{t \rightarrow 1} \left( P^{T[2,2p+1]}_{\Lambda^r} \right)
(-1)^rq^{r(2-p)}A^{2r(p+1)}\frac{(1-A^{-2}q^{-2})^r}{(1-q^2)^r}
\sum\limits^r_{l=0} \frac
{(1-A^{-2})^l}
{(1-A^{-2}q^{-2})^l}
q^{2pl}C^r_l
\ee
Since $\ \sum\limits^r_{l=0} \frac
{(1-A^{-2})^l}
{(1-A^{-2}q^{-2})^l}
q^{2pl}C^r_l
=\left(\frac{A^2-1}{A^2-q^{-2}}q^{2p}\right)^r$ we obtain:
\be
\boxed{{\frak P}^{T[2,2p+1]}_{\Lambda^r}= \lim_{t\rightarrow1}P^{T[2,2p+1]}_{\Lambda^r} (-1)^rq^{r(2-p)}A^{2r(p+1)}\frac{(A^2-q^{-2})^r}{(A^2-q^2)^r}
(\frac{A^2-1}{A^2-q^{-2}}q^{2p})^r=\left({\frak P}^{T[2,2p+1]}_{\Lambda^1}\right)^r}
\label{asts}
\ee
So our conjecture (\ref{3}) in the antisymmetric case is correct for generic $p$ and $r$ too!
\section{The case of figure-eight knot}
The only non-torus case when colored superpolynomials are known for the entire series of representations is the figure-eight knot $4_1$. According to \cite{IMMM} we have the following expression for reduced symmetric figure-eight superpolynomials:
\be
  {P}^{4_1}_{[S^r]}(A,t,q)=\sum\limits_{k=0}^r  \sum_{1\leq i_1<...<i_k\leq p} {\frak Z}_{i_1}(A) {\frak Z}_{i_2}(Aq) {\frak Z}_{i_k}(Aq^{k-1})
\ee
where
\be
{\frak Z}_i(Aq^k) =\big\{A q^{2(p-i)+1+k}\big\}\left\{\frac{Aq^k}{t}\right\}
\ee
and $\{x\}=x-\frac{1}{x}$

In the case of $q\rightarrow1$ all the $\frak{Z}$ factors in (12) are the same and superpolynomial becomes:
\be
{\cal P}^{4_1}_{[S^r]}(A,t)=(1+{\frak Z})^r=\left({\cal P}^{T[4_1]}_{[1]}(A,t)\right)^r
\ee

Similarly for antisymmetric case:
\be
  {P}^{4_1}_{[\Lambda^r]}(A,t,q)=\sum\limits_{k=0}^r  \sum_{1\leq i_1<...<i_k\leq p} \overline{{\frak Z}}_{i_1}(A) \overline{{\frak Z}}_{i_2}(Aq) \overline{{\frak Z} }_{i_k}(Aq^{k-1})
\ee
where
\be
\overline{{\frak Z}}_i(At^{-k}) =\left\{\frac{A}{t^{2(p-i)+s+1}}\right\}\left\{\frac{qA}{t^k}\right\}
\ee
In the case of $t\rightarrow1$ it becomes:
\be
{\frak P}^{4_1}_{[\Lambda^r]}(A,q)=(1+\overline{{\frak Z}})^r=\left({\frak P}^{4_1}_{[1]}(A,q)\right)^r
\ee
Thus for the figure-eight knot our conjectures (\ref{2}) and (\ref{3}) are also true.

In fact the figure-eight case is somewhat special because our colored superpolynomials satisfy the extraordinary simple recursion relations. According to \cite{IMMM} formulas $(12)$ and $(15)$ imply that
\be
P_{[S^{r+1}]} - P_{[S^r]}(A)= \{At^{-1}\}\{A q^{2r+1}\}P_{[S^r]} (Aq)
\ee

\be
P_{[\Lambda^{r+1}]} - P_{[\Lambda^r]}(A)= \{At^{-1-2r}\}\{A q\}P_{[\Lambda^r]} (At^{-1})
\ee
for the symmetric and the antisymmetric representations respectively. When $q=1$ the superpolynomial at the right hand side of $(18)$ is the same as the second item at the left hand side and we get from the recursion relation a product and even a {\it power} formula $(14)$. To get the same from $(19)$ we need to put $t=1$.
It is unclear if the product formulas exist also for non-hook and even for hook representations.
\section{Conclusion}
To conclude, we considered some cases of {\it reduced} colored superpolynomials and observed an interesting dependence on representation $R$, which generalizes the property (\ref{4}) of special HOMFLY polynomials. The difference is that (\ref{4}) is proved for arbitrary knots and even links, and is true for arbitrary representation $R$. Our (\ref{2}) and (\ref{3}) are only conjectures: we believe that they hold for arbitrary knots but actually checked this only for 2-strand knots and figure-eight. Most important, the dependence on representation $R$ is not as universal as it was in (\ref{4}): for totaly symmetric $R$ we have (\ref{2}) while for totaly antisymmetric - (\ref{3}), with different (but related) reduced superpolynomials, and  it is a bit mysterious what happens for non-special representations. Instead, (\ref{2}) and (\ref{3}) are formulas for families of polynomials, depending on two parameters $(A,t)$ and $(A,q)$, while (\ref{4}) depends only on one $(A)$.

The main problem with investigation of our conjecture is the lack  of examples of colored superpolynomials. From this point of view, it is extremely important to find at least some superpolynomials for the simplest hook and non-hook diagrams $R=[2,1]$ and $R=[2,2]$.

\section*{Acknowledgements}
I am indebted to Alexei Sleptsov for formulating the problem
and to Andrey Morozov for numerous explanations.
My work is partly supported by the
Ministry of Education and Science of the Russian Federation under the contract
2012-1.1-12-000-1003-032 (12.740.11.0677),
by the grants
NSh-14.740.11.0347,
RFBR-12-01-00525 and
the Young Researcher grant 12-02-31280.

\end{document}